\documentclass[preprint,showpacs,showkeys,aps,preprintnumbers,amsmath,amssymb]{revtex4}
 \usepackage{dcolumn}
 \usepackage{bm}

\begin{document}

 \title{``Charged'' Particle' s Tunneling from Rotating Black Holes}

\author{Tao Zhu}

\email{zhut05@gmail.com}

\affiliation{Department of Physics, Zhejiang University of Technology, Hangzhou 310032, China\\}

 \begin{abstract}

 The behavior of a scalar field theory near the event horizon in a rotating black hole background can be effectively described by a two dimensional field theory in a gauge field background. Based on this fact, we proposal that the quantum tunneling from rotating black hole can be treated as ``charged'' particle' s tunneling process in its effectively two dimensional metric. Using this viewpoint and considering the corresponding ``gauge charge'' conservation, we calculate the non-thermal tunneling rate of Kerr black hole and Myers-Perry black hole, and results are consistent with Parikh-Wilczek' s original result for spherically symmetric black holes. Especially for Myers-Perry black hole which has multi-rotation parameters, our calculation fills in the gap existing in the literature applying Parikh-Wilczek' s tunneling method to various types black holes. Our derivation further illuminates the essential role of effective gauge symmetry in
Hawking radiation from rotating black holes.

 \end{abstract}

 \maketitle

\tableofcontents

\def\thesection{\arabic{section}}
\def\thesubsection{\arabic{section}.\arabic{subsection}}
\numberwithin{equation}{section}

\section{Introduction}
\label{INT}

Hawking radiation arises upon the quantum effect of matter in a classical background spacetime with an event horizon.
It therefore plays an important role in black hole physics.  Apart from Hawking' s original derivation\cite{CMP}, there
are many other approaches in literature. In 2000,  Parikh and Wilczek put forward a semiclassical framework to implement
Hawking radiation as a tunneling process across the horizon of the static spherically symmetric black holes and
demonstrated that the emission spectrum of black hole radiance is
not strictly pure thermal if the energy conservation (or self-gravitation) is considered\cite{PRL2000}. Their results are considered to be in agreement with an
underlying unitary theory, and thus could provide insight into the information loss paradox\cite{information,JHEP2010}. Following Parikh-Wilczek' s tunneling
method, various spherically symmetric black holes are studied\cite{sbhs}, and all the obtained results are
very successful to support the Parikh-Wilczek' s prescription. Parikh-Wilczek' s original calculation only considered the tunneling process of a massless and uncharged particle, recently it is shown that such tunneling method can be easily extended to study the massive\cite{massive} and charged particle's tunneling process\cite{charge}. For the charged particle's tunneling, both the energy conservation and charge conservation should be taken into account.

There also have a lot of attempts to extend the tunneling approach of Hawking radiation to the case of some stationary rotating black hole backgrounds\cite{rotating black hole}. There are two key tricks to calculate the tunneling rate of rotating black hole. First, due to the rotation degree of freedom, the tunneling particle might contain angular momentum. Thus, in order to investigate the tunneling process from a rotating black hole, not only the energy conservation but also the angular momentum conservation should be considered. Second, one must consider the dragging effect in a rotating black hole spacetime
since the physical field must be dragged also in a rotating background spacetime. Thus, the description of the tunneling process must be within a dragging coordinate system. For various rotating black hole with only one rotation parameter, their dragging coordinate systems can be well defined and thus Parikh-Wilczek' s tunneling method has been successful applied. Other recent work in this direction can be found in \cite{chow,time}

However, when we considered the high dimensional rotating black hole with more than one rotation parameters, things becomes more complicated. In such case, a dragging coordinate system which can eliminate all the rotation degree of freedoms, usually cannot be found. Due to this difficulty, up to now, most of the studies
are focus on the rotating black holes with only one rotation. To see the universality of Hawking radiation, it is meaningful to extend Parikh-Wilczek' s tunneling method to more general black hole with more than one rotational parameters.

On the other hand, it is known that a $(n+1)$-dimensional black hole metric effectively becomes a $2$-dimensional spherically symmetric metric by using the technique of the dimensional reduction near the horizon\cite{dimensional reduce,Kerr,MP}. Such effective $2$-dimensional metric has a very simple form, and thus Parikh-Wilczek' s tunneling
method can be easily used\cite{dtunneling}. For rotating black hole, each partial wave of quantum fields in black hole background can be interpreted as a $(1+1)$-dimensional charged field in an effective gauge field background. In this case, a rotation degree of freedom effectively becomes a gauge degree of freedom. Multi-rotations mean multi-gauge fields. This feature leads to a interesting question. That is, when one investigate the tunneling process in rotating background, the conservation of angular momentum should be considered, now for its effective $2$-dimensional metric, how can one apply the tunneling method? This question needs answer. Furthermore, it has been argued that both entropy and Hawking radiation of black holes are deeply connected to the conformal symmetry arising near the horizon\cite{dimensional reduce,Kerr,Hawking-Cft}. For rotating black holes, its effective gauge symmetry near the horizon is at the heat of connections between conformal symmetry and black hole radiation\cite{Kerr,Hawking-Cft}. Thus, investigating Hawking radiation from the effective two dimensional metric near the horizon not only shows the universality of Hawking radiation, but also might shed the light on holographic dual of black hole physics and conformal field theory.

Combine above reasons, in this paper we are going to investigate the Hawking radiation as tunneling process form two rotating black holes by using the dimensional reduction technique. We consider two types of rotating black holes, which are $4$-dimensional Kerr black hole and $D$-dimensional Myers-Perry black hole with multi-rotations. It is shown that using their effective $2$-dimensional metric from dimensional reduction, the tunneling process of the particle with angular momentum can be treated as the ``charged'' particle's tunneling process and the obtained results support the Parikh-Wilczek' s prescription. Thus this investigation fills in the gap existing in the literature applying Parikh-Wilczek' s tunneling method to various types black holes. Our investigation emphasizes the essential role of effective gauge symmetry in derivation of tunneling rate of rotating black holes, which may suggest new insight
on the understanding of black hole radiance.

This paper is organized as follows: In section 2, by using the effective two dimensional metric from dimensional reduction of a scalar field action in Kerr black hole background, we consider the ``charged'' particle's tunneling from this two dimensional metric and calculate its tunneling rate. In section 3, we extend this analysis to Myers-Perry black hole. The
last section is devoted to conclusions.

\section{``Charged'' particle's tunneling from Kerr black hole}
In this section, we consider the quantum tunneling from Kerr black hole, whose
line element can be
written in the form
\begin{eqnarray}
ds^2&=&-\frac{\Delta}{r^2+a^2\cos^2{\theta}}\left(dt-a\sin^2{\theta}d\varphi\right)^2+\frac{\sin^2{\theta}}{r^2+a^2\cos^2{\theta}}\left(adt-(r^2+a^2)d\varphi\right)^2\nonumber\\
&&+\left(r^2+a^2\cos^2{\theta}\right)\left(\frac{dr^2}{\Delta}+d\theta^2\right),
\end{eqnarray}
where
\begin{eqnarray}
\Delta=r^2-2Mr+a^2=(r-r_{+})(r-r_{-}),
\end{eqnarray}
and $r_{+(-)}$ are radius of outer (inner) horizons
\begin{eqnarray}
r_{\pm}=M\pm\sqrt{M^2-a^2}.
\end{eqnarray}

Now we begin to apply the technique of the dimensional reduction near the horizon of Kerr black hole. For simplicity, we consider a scalar field in the Kerr black hole background, whose action is
\begin{eqnarray}
S &=& \frac{1}{2}\int d^4x\sqrt{-g}g^{\mu\nu}\partial_\mu\phi\partial_\nu\phi+S_{int}\nonumber\\
&=& \left.\frac{1}{2}\int dtdrd\theta d\varphi\sin{\theta}\phi\right[\left(\frac{(r^2+a^2)^2}{\Delta}-a^2\sin^2{\theta}\right)\partial_t^2+
2a\left(\frac{r^2+a^2}{\Delta}-1\right)\partial_t\partial_\varphi\nonumber\\
&& -\partial_r\Delta\partial_r-\frac{1}{\sin^2{\theta}}\partial_\theta\sin^2{\theta}\partial_\theta-\left.
\frac{1}{\sin^2{\theta}}\left(1-\frac{a^2\sin^2{\theta}}{\Delta}\right)\partial^2_\varphi\right]\phi+S_{int},
\end{eqnarray}
where $S_{int}$ represents the mass, potential and interaction terms. By performing the partial wave decomposition of $\phi$ in terms of the spherical harmonics
\begin{eqnarray}
\phi=\sum_{l,m}\phi_{lm}(t,r)Y_{lm}(\theta,\varphi),
\end{eqnarray}
the theory is reduced to a two-dimensional effective theory with an infinite collection of fields with quantum numbers $(l,m)$.
Upon transforming to the $r_*$ tortoise coordinate, defined by
\begin{eqnarray}
\frac{dr_*}{dr}=\frac{r^2+a^2}{\Delta}=f(r)^{-1},
\end{eqnarray}
and considering the region near the outer horizon $r_+$, one finds that the effective radial potentials for partial waves or mixing terms between fields with different angular momenta contain a suppression factor and vanish exponentially fast near the horizon. Also we can ignore all the terms in $S_{int}$ due to the same reason. After this analysis, one can obtain\cite{Kerr}
\begin{eqnarray}
S=\int dtdr(r^2+a^2)\phi_{lm}^*\left[\frac{r^2+a^2}{\Delta}\left(\partial_t+\frac{iam}{r^2+a^2}\right)^2-\partial_r\frac{\Delta}{r^2+a^2}\partial_r\right]\phi_{lm}.\label{2action}
\end{eqnarray}
From above one finds that $\phi_{lm}$ can be considered as a $(1+1)$-dimensional complex scalar field in the backgrounds of the dilaton $\Phi$, metric $g_{\mu\nu}$ and $U(1)$ gauge field $A_\mu$,
\begin{eqnarray}
\Phi&=&r^2+a^2,\\
g_{tt}&=&-f(r),~~~g_{rr}=f(r)^{-1}, ~~g_{tr}=g_{rt}=0,\\
A_t&=&-\frac{a}{r^2+a^2},~~~~~A_r=0.
\end{eqnarray}
The $U(1)$ gauge charge of the two dimensional field $\phi_{lm}$ is $m$.

From the action (\ref{2action}) we can see that the $4$-dimensional Kerr metric behaves as the $2$-dimensional spherically symmetric line element in the region near the horizon
\begin{eqnarray}
ds_2^2=-f(r)dt^2+\frac{1}{f(r)}dr^2.
\end{eqnarray}
Because the tunneling method deals with the region very close to
the horizon, one can investigate the quantum tunneling effect of Kerr black hole by using this two dimensional metric\cite{dtunneling}.
To describe the across-horizon phenomena,  we should introduce the Painlev\'e-like coordinate system by the transformation
\begin{eqnarray}
dt_k=dt+\frac{\sqrt{2Mr}}{1-\frac{2Mr}{r^2+a^2}}dr,
\end{eqnarray}
then the Painlev\'e-like coordinate is given by
\begin{eqnarray}
ds^2=-\left(1-\frac{2Mr}{r^2+a^2}\right)dt_k^2+2\sqrt{\frac{2Mr}{r^2+a^2}}dt_kdr+dr^2.
\end{eqnarray}
The radial null geodesics are given by
\begin{eqnarray}\label{null}
\dot{r}=\pm 1-\sqrt{\frac{2Mr}{r^2+a^2}},
\end{eqnarray}
with the upper(lower) sign corresponding to outgoing (ingoing) geodesics.

Since we are considering a rotating black hole, so the
rotation degree of freedom should be well addressed also.
Thus, when we consider the tunneling process near the horizon, both the energy conservation and angular momentum
conservation should be taken into account. From action (\ref{2action}) we can see that only
the magnetic quantum number $m$ of particle is relevant near the horizon, and the particle which contains quantum number $m$ behaves as a ``charged'' particle with gauge charge $m$ in the background gauge field $A_\mu$. In this picture, the angular momentum conservation means gauge charge $m$ conservation. Thus we can treat the tunneling process as the ``charged'' particle's tunneling. Moreover, if the self-gravitation of the tunneling particle is included, eq.(\ref{null}) should be replaced by $M\rightarrow M-\omega$, $a=\frac{J}{M}\rightarrow a'=\frac{J-m}{M-\omega}$, thus
\begin{eqnarray}\label{null2}
\dot{r}=\pm 1-\sqrt{\frac{2(M-\omega)r}{r^2+a'^2}},
\end{eqnarray}
where $\omega$ is the particle energy and $J$ is the total angular momentum of Kerr black hole.

When we investigate a charged particle' s tunneling process, the effect of the gauge field
should be taken into account. To include the gauge field effect, we write the lagrangian
function of the system as $L=L_A+L_m$, where $L_A=-F_{\mu\nu}F^{\mu\nu}/4$ is the lagrangian
function of the gauge field corresponding to the generalized coordinates $A_\mu=(-\frac{a}{r^2+a^2},0)$. When a charged particle tunnels
out, the system transit from one state to another.  But from the expression of $L_A$ we find that $A_t$ is an ignorable coordinate. To eliminate the freedom corresponding to $A_t$, the action should be written as
\begin{eqnarray}
I=\int_{t_i}^{t_f}(L-P_{A_t}\dot{A}_t)dt,
\end{eqnarray}
which is related to the emission rate of the tunneling particle by
\begin{eqnarray}
\Gamma\sim e^{-2\texttt{Im} I}.
\end{eqnarray}
Therefore, the imaginary part of the action is
\begin{eqnarray}\label{IP}
\texttt{Im} I=\texttt{Im}\int_{t_i}^{t^f}\left[P_r-\frac{P_{A_t}\dot{A}_t}{\dot{r}}\right]dr=
\texttt{Im}\int_{(0,0)}^{(P_r,P_{A_t})}\left[dP'_r-\frac{\dot{A}_t}{\dot{r}}dP'_{A_t}\right]dr,
\end{eqnarray}
where $P_{A_t}$ is the gauge field's canonical momentum conjugate to $A_t$. To proceed with an explicit calculation,
we now remove the momentum in favor of energy by applying the Hamilton's equations
\begin{eqnarray}
\label{H1}\dot{r}&=&\left.\frac{dH}{dP_r}\right|_{(r;A_t,P_{A_t})}=\frac{d(M-\omega')}{dP_r},\\
\label{H2}\dot{A}_t&=&\left.\frac{dH}{dP_{A_t}}\right|_{(A_t;r,P_r)}=-\frac{a'}{r^2+a'^{2}}\frac{dm'}{dP_{A_t}},
\end{eqnarray}
where $a'=\frac{J-m'}{M-\omega'}$ and $dH_{(A_t;r,P_r)}$ represents the energy change of the black hole because of the loss of the gauge charge $m$ when a particle tunnels out.

Substituting Eqs. (\ref{null2}), (\ref{H1}) and (\ref{H2}) into Eq. (\ref{IP}), and noting
that we must choose the positive sign in Eq. (\ref{null2}) as the
particle is propagating from inside to outside the event horizon, then we have
\begin{eqnarray}
\texttt{Im} I=\texttt{Im}\int_{(M,0)}^{M-\omega,m}\int_{r_i}^{r_f}\left[d(M-\omega')+\frac{a'}{r^2+a'^2}dm'\right]\frac{dr}{1-\sqrt{\frac{2(M-\omega')r}{r^2+a'^2}}}.
\end{eqnarray}
We see that $r=r'_+=M-\omega'+\sqrt{(M-\omega')^2-a'^2}$ is a single pole in above equation.  The integral can be evaluated by deforming
the contour around the pole. In this way we obtain
\begin{eqnarray}
\texttt{Im} I
=-2\pi\int_{(M,J)}^{(M-\omega,J-m)}\frac{r'^{2}_++a'^2}{r'_+-r'_{-}}\left[d(M-\omega')-\frac{a'}{r'^2_{+}+a'^2}d(J-m')\right],
\end{eqnarray}
where $r_-=M-\omega'-\sqrt{(M-\omega')^2-a'^2}$. In above we have used $dm'=-d(J-m')$. Notice that the Hawking temperature on the horizon of Kerr black hole is
\begin{eqnarray}
T'=\frac{r'_+-r'_{-}}{4\pi(r'^{2}_++a'^2)}.
\end{eqnarray}
Thus we have
\begin{eqnarray}
\texttt{Im} I=-\frac{1}{2}\int_{(M,J)}^{(M-\omega,J-m)}\frac{1}{T'}\left[d(M-\omega')-\frac{a'}{r'^2_{+}+a'^2}d(J-m')\right]=-\frac{1}{2}\Delta S_{\texttt{BH}},
\end{eqnarray}
where $\Delta S_{\texttt{BH}}$ is the difference of the entropies of the
black hole before and after the emission. The tunneling rate is therefore
\begin{eqnarray}
\Gamma\sim e^{-2\texttt{Im}I}=e^{\Delta S_{\texttt{BH}}},
\end{eqnarray}
which is consistent with the result in \cite{PRL2000} obtained from spherically symmetric black holes.

\section{``Charged'' particle' s tunneling from Myers-Perry black hole}

The Kerr black hole in $4$-dimensional spacetime only has one rotation, in this section, we turn to consider the quantum tunneling from Myers-Perry black hole which has more than one rotation parameters. In the $D=2n+1+\epsilon ~(\epsilon = 0 \mbox{ or } 1)$ dimensions,
the metric of the Myers-Perry black hole is given by \cite{MP}
\begin{eqnarray}
 ds^2 &=& -dt^2 + \epsilon r^2 d\alpha^2
  + \sum_{i=1}^n (r^2+a_i^2)
  \left(d\mu_i^2+\mu_i^2d\phi_i^2\right)
  + \frac{\mu r^{2-\epsilon}}{\Pi F}
  \left(dt - \sum_{i=1}^n a_i \mu_i^2d\phi_i\right)^2 \nonumber \\
  && + \frac{\Pi F}{\Pi - \mu r^{2-\epsilon}}dr^2,
 \end{eqnarray}
where
\begin{eqnarray}
 F &=& 1-\sum_{i=1}^n \frac{a_i^2\mu_i^2}{r^2+a_i^2}, \\
 \Pi &=& \prod_{i=1}^n (r^2 + a_i^2).
\end{eqnarray}
The following constraint is imposed for $\mu_i ~(i=1, 2, \cdots,n)$ and
$\alpha$,
\begin{eqnarray}
 \sum_{i=1}^n \mu_i^2 + \epsilon\alpha^2 = 1,
  \qquad (0\leq \mu_i \leq 1, ~-1 \leq \alpha \leq 1).
\end{eqnarray}
This metric describes a black hole spacetime with the mass
$M=\frac{(D-2)A_{D-2}}{16\pi G}\mu$ and angular momenta $\frac{2}{D-2} M
a_i$ in the $\phi_i$-directions, where $A_{D-2}$ is the volume of $S^{D-2}$.
This black hole is stationary and has $U(1)^n$ isometries with the Killing
vectors $\partial_{\phi_i}$.  We assume the existence of horizons located at
positive solutions of $\Pi - \mu r^{2-\epsilon}=0$.

The action of a scalar field in Myers-Perry background is
\begin{eqnarray}
S=\frac{1}{2}\int d^4x\sqrt{-g}g^{\mu\nu}\partial_\mu\phi\partial_\nu\phi+S_{int}
\end{eqnarray}
By performing the expansion of scalar field $\phi$ by the $Y_{m_1\dots m_n}(\mu_i,\phi_i)$, which is the spherical harmonics on $S^{D-1}$, i.e.,
\begin{eqnarray}
\phi=\sum_{m_i}\phi_{m_1\dots m_n}(t,r)Y_{m_1\dots m_n}(\mu_i,\phi_i),
\end{eqnarray}
then the physics of scalar field near the horizon can be effectively described by an infinite collection of massless $(1+1)$-dimensional fields with following action,
\begin{eqnarray}\label{MPaction}
S=\int dtdr(\mu r)\phi^{*}_{m_1\dots m_n}\left[\frac{\Pi}{\Pi-\mu r^{2-\epsilon}}\left(\partial_t+\sum_{i=1}^{n}\frac{im_ia_i}{r^2+a_i^2}\right)-\partial_r\frac{\Pi-\mu r^{2-\epsilon}}{\Pi}\partial_r\right]\phi_{m_1\dots m_n}.\nonumber\\
\end{eqnarray}

From this action we find that $\phi_{m_1\dots m_n}$ can be considered as a $(1+1)$-dimensional
complex scalar field in the background of the dilaton $\Phi$, metric $g_{mu\nu}$ and $U(1)$ gauge fields
$A_\mu^{(\phi_i)}$,
\begin{eqnarray}
\Phi&=&\mu r,\nonumber\\
g_{tt}&=&-\frac{\Pi-\mu r^{2-\epsilon}}{\Pi}=-f(r),~~~~g_{rr}=\frac{\Pi}{\Pi-\mu r^{2-\epsilon}}=\frac{1}{f(r)},~~~~g_{tr}=0,\nonumber\\
A_t^{(\phi_i)}&=&-\frac{a_i}{r^2+a_i^2}, ~~~~A_r^{(\phi_i)}=0.
\end{eqnarray}
The partial wave $\phi_{m_1\dots m_n}$ has the $U(1)$ charge $m_i$ for each gauge field $A_{\mu}^{(\phi_i)}$.

From the action (\ref{2action}) we can see that the $D$-dimensional Myers-Perry metric behaves as the $2$-dimensional spherically symmetric line element in the region near the horizon
\begin{eqnarray}
ds_2^2=-f(r)dt^2+\frac{1}{f(r)}dr^2.
\end{eqnarray}
Because the tunneling method deals with the region very close to
the horizon, one can investigate the quantum tunneling effect of Myers-Perry black hole by using this two dimensional metric.
To describe the across-horizon phenomena,  we should introduce the Painlev\'e-like coordinate system by the transformation
\begin{eqnarray}
dt_k=dt+\frac{\sqrt{\Pi \mu r^{2-\epsilon}}}{\Pi-\mu r^{2-\epsilon}}dr,
\end{eqnarray}
then the Painlev\'e-like coordinate is given by
\begin{eqnarray}
ds^2=-\left(\frac{\Pi-\mu r^{2-\epsilon}}{\Pi}\right)dt_k^2+2\sqrt{\frac{\mu r^{2-\epsilon}}{\pi}}dt_kdr+dr^2.
\end{eqnarray}
The radial null geodesics are given by
\begin{eqnarray}\label{mpnull}
\dot{r}=\pm 1-\sqrt{\frac{\mu r^{2-\epsilon}}{\Pi}},
\end{eqnarray}
with the upper(lower) sign in Eq. (\ref{null}) corresponding to outgoing (ingoing) geodesics.

For the tunneling process in Myers-Perry background, both the energy conservation and angular momentum
conservation should be taken into account. From action (\ref{MPaction}) we can see that only
the magnetic quantum number $m_i$ of particle is relevant near the horizon, and the particle which contains quantum number $m_i$ behaves as a ``charged" particle with gauge charge $m_i$ in the background gauge field $A^{(\phi_i)}_\mu$. In this picture, the angular momentum conservation means gauge charge $m_i$ conservation. Thus we can treat the tunneling process as the ``charged" particle's tunneling. Unlike Kerr black hole, here the tunneling particle can contains multi gauge charges.

To include the gauge field effect in the tunneling process, we write the lagrangian
function of the system as $L=\sum_{i}L_{A^{(\phi_i)}}+L_m$, where $L_{A^{(\phi_i)}}=-F^{(\phi_i)}_{\mu\nu}F^{(\phi_i)\mu\nu}/4$ is the lagrangian
function of the gauge field corresponding to the generalized coordinates $A^{(\phi_i)}_\mu=(-\frac{a_i}{r^2+a_i^2},0)$. When a charged particle tunnels
out, the system transit from one state to another.  But from the expression of $L_{A^{(\phi_i)}}$ we find that $A^{(\phi_i)}_t$ is an ignorable coordinate. To eliminate the freedom corresponding to $A^{(\phi_i)}_t$, the action should be written as
\begin{eqnarray}
I=\int_{t_i}^{t_f}(L-\sum_{i}P_{A^{(\phi_i)}_t}\dot{A}^{(\phi_i)}_t)dt,
\end{eqnarray}
the imaginary part of the action is written as
\begin{eqnarray}\label{mpIP}
\texttt{Im} I&=&\texttt{Im}\int_{t_i}^{t_f}\left[P_r-\sum_i\frac{P_{A^{(\phi_i)}_t}\dot{A}^{(\phi_i)}_t}{\dot{r}}\right]dr\nonumber\\
&=&\texttt{Im}\int_{(0,0)}^{(P_r,P_{A^{(\phi_i)}_t})}\left[dP'_r-\sum_i\frac{\dot{A}^{(\phi_i)}_t}{\dot{r}}dP'_{A^{(\phi_i)}_t}\right]dr,
\end{eqnarray}
where $P_{A^{(\phi_i)}_t}$ is the gauge field's canonical momentum conjugate to $A^{(\phi_i)}_t$. To proceed with an explicit calculation,
we now remove the momentum in favor of energy by applying the Hamilton's equations
\begin{eqnarray}
\label{mp1}\dot{r}&=&\left.\frac{dH}{dP_r}\right|_{(r;A^{(\phi_i)}_t,P_{A^{(\phi_i)}_t})}=\frac{d(M-\omega')}{dP_r},\\
\label{mp2}\dot{A}^{(\phi_i)}_t&=&\left.\frac{dH}{dP_{A^{(\phi_i)}_t}}\right|_{(A^{(\phi_i)}_t;r,P_r)}=-\frac{a^{'}_{i}}{r^2+a_{i}^{'2}}\frac{dm^{'}_i}{dP_{A^{(\phi_i)}_t}},
\end{eqnarray}
where $a^{'}=\frac{J-m^{'}}{M-\omega^{'}}$ and $dH_{(A^{(\phi_i)}_t;r,P_r)}$ represents the energy change of the black hole because of the loss of the gauge charge $m_i$ when a particle tunnels out.

Combining Eqs. (\ref{mpnull}), (\ref{mp1}) and (\ref{mp2}), and noting
that we must choose the positive sign in Eq. (\ref{mpnull}) as the
particle is propagating from inside to outside the event horizon, then Eq. (\ref{mpIP}) can be expressed as
\begin{eqnarray}
\texttt{Im} I=\texttt{Im}\int_{(M,0)}^{(M-\omega,m_i)}\int_{r_i}^{r_f}\left[d(M-\omega')+\sum_i\frac{a^{'}_{i}}{r^2+a^{'2}_i}dm'\right]\frac{dr}{1-\sqrt{\frac{\mu^{'} r^{2-\epsilon}}{\Pi^{'}}}}.
\end{eqnarray}
We see that there is a single pole located at the horizon $r=r^{'}_{+}$.  The integral can be evaluated by deforming
the contour around the pole. In this way we obtain
\begin{eqnarray}
\texttt{Im} I=-\int_{(M,J)}^{(M-\omega,J_i-m_i)}\frac{2\pi}{f'(r_+)}\left[d(M-\omega')-\sum_i\frac{a_i^{'}}{r'^2_{+}+a_i^{'2}}d(J_i-m_i^{'})\right],
\end{eqnarray}
where $f'(r_+)=\left.\frac{df(r)}{dr}\right.|_{r=r_+}$. In above we have used $dm_i^{'}=-d(J_i-m_i^{'})$. Notice that the Hawking temperature on the horizon of Kerr black hole is
\begin{eqnarray}
T'=\frac{f'(r_+)}{4\pi}.
\end{eqnarray}
Thus we have
\begin{eqnarray}
\texttt{Im} I&=&-\frac{1}{2}\int_{(M,J_i)}^{(M-\omega,J_i-m_i)}\frac{1}{T'}\left[d(M-\omega')-\sum_i\frac{a_i^{'}}{r'^2_{+}+a_i^{'2}}d(J_i-m_i^{'})\right]\nonumber\\
&=&-\frac{1}{2}\Delta S_{\texttt{BH}},
\end{eqnarray}
where $\Delta S_{\texttt{BH}}$ is the difference of the entropies of the
black hole before and after the emission. The tunneling rate is therefore
\begin{eqnarray}
\Gamma\sim e^{-2\texttt{Im}I}=e^{\Delta S_{\texttt{BH}}},
\end{eqnarray}
which is consistent with the result in \cite{PRL2000} obtained from spherically symmetric black holes.

\section{Conclusions and Discussions}

In this paper, the dimensional reduction technique is applied to investigate the quantum tunneling process from rotating black hole backgrounds. We have found that the quantum tunneling from rotating black holes can be treated as the ``charged'' particle' s tunneling process in the reduced two dimensional spacetime background. This is the essential observation in this paper. With this essential viewpoint, we have shown that Parikh-Wilczek' s tunneling method can be successfully applied to two types of rotating black holes, which are Kerr black hole in $4$ dimensional spacetime and Myers-Perry black hole with multi-rotation parameters in $D$ dimensional spacetime. The obtained tunneling rates are non-thermal and consistent with Parikh-Wilczek's original result for spherically symmetric black holes. Especially for Myers-Perry black hole, our calculation fills in the gap existing in the literature applying Parikh-Wilczek's tunneling method to various types black holes.

On the other hand, the dimensional reduction shows that near the horizon, except magnetic quantum number, any physical scale of the tunneling particle is sweeps away due to the exponential blueshift associated with the event horizon. This magnetic quantum number behaves as the conservation charge of the effective U(1) gauge symmetry near the horizon. The manifestation of such effective gauge symmetry has played an essential role in studying Hawking radiation from rotating black holes. In quantum anomaly method, the Hawking flux of the thermal radiation from rotating black hole is deeply connected to anomalies of such effective gauge symmetry\cite{Kerr,MP}, and in this paper we applied this gauge symmetry in quantum tunneling method. Our study here further shows that the manifestation of the effective gauge symmetry near the horizon in rotating black hole is deeply related to Hawking radiance. It provides a new perspective for understanding black hole radiance and might shed the light on the connection between the conformal symmetry and
black hole radiance\cite{Hawking-Cft}.


 \section*{Acknowledgement}


\begin{thebibliography}{99}
 \bibitem{CMP}
 S.W. Hawking, %
 {\em Particle creation by black holes}, %
 Commun. Math. Phys. \textbf{43} (1975) 199. %

 \bibitem{PRL2000}
 M.K. Parikh and F. Wilczek, %
 {\em Hawking Radiation As Tunneling}, %
 Phys. Rev. Lett. \textbf{85} (2000) 5042 [hep-th/9907001].

 \bibitem{JHEP2010}
 D. Singleton, E.C. Vagenas, T. Zhu, and J-R. Ren, %
{\em Insights and possible resolution to the information loss paradox via the tunneling picture}, %
 JHEP \textbf{08} (2010) 089 [arXiv:1005.3778[hep-th]].

 \bibitem{information}
 M.K. Parikh, %
{\em A Secret Tunnel Through The Horizon}, %
 Int. J. Mod. Phys. \textbf{D 13} (2004) 2351 [hep-th/0405160]. %
 \\
 M. Arzano, A.J.M. Medved, and E.C. Vagenas, %
{\em Hawking Radiation as Tunneling through the Quantum Horizon}, %
 JHEP \textbf{09} (2005) 037 [hep-th/0505266]. %
 \\
 B. Zhang, Q-yu Cai, L. You, and M.S Zhan, %
{\em Hidden Messenger Revealed in Hawking Radiation: a Resolution to the Paradox of Black Hole Information Loss}, %
 Phys. Lett. \textbf{B 675} (2009) 98 [arXiv:0903.0893[hep-th]]; %
 \\
 B. Zhang, Q-yu Cai, M-s Zhan, and L. You, %
{\em No Information Is Lost: a Revisit of Black Hole Information Loss Paradox}, %
 arXiv:0906.5033[hep-th]. %
 \\
 Y-X. Chen and K-N. Shao, %
{\em Information loss and entropy conservation in quantum corrected Hawking radiation}, %
 Phys. Lett. \textbf{B 678} (2009) 131 [arXiv:0905.0948[hep-th]].

 \bibitem{sbhs}
 M.K. Parikh, %
{\em New coordinates for de Sitter space and de Sitter radiation}, %
 Phys. Lett. \textbf{B 546} (2002) 189 [hep-th/0204107]; %
{\em Energy conservation and Hawking radiation}, %
hep-th/0402166; %
\\
S. Hemming and E. K. Vakkuri, %
{\em Hawking radiation from AdS black holes}, %
Phys. Rev. \textbf{D 64} (2001) 044006 [gr-qc/0005115]; %
\\
A.J.M. Medved, %
{\em Radiation via tunneling from a de Sitter cosmological horizon}, %
Phys. Rev. \textbf{D 66} (2002) 124009 [hep-th/0207247]; %
\\
E.C. Vagenas, %
{\em Generalization of the KKW analysis for black hole radiation}, %
Phys. Lett. \textbf{B 559} (2003) 65 [hep-th/0209185]; %
\\
M. Arzano, %
{\em Tunneling through the quantum horizon}, %
Mod. Phys. Lett. \textbf{A 21} (2006) 41 [hep-th/0504188]; %
\\
A.J.M. Medved and E.C. Vagenas, %
{\em On Hawking radiation as tunneling with back-reaction}, %
Mod. Phys. Lett. \textbf{A 20} (2005) 2449 [gr-qc/0504113]; %
\\
I. Radinschi, %
{\em The KKW generalized analysis for a magnetic stringy black hole}, %
gr-qc/0412111.


\bibitem{massive}
J. Zhang and Z. Zhao, %
{\em Massive particles' black hole tunneling and de Sitter tunneling}, %
Nucl. Phys. \textbf{B 725} (2005) 173.

\bibitem{charge}
J. Zhang and Z. Zhao, %
{\em Hawking radiation of charged particles via tunneling from the Reissner-Nordstrom black hole}, %
JHEP \textbf{10} (2005) 055; %
\\
Q.-Q. Jiang and S.-Q. Wu, %
{\em Hawking radiation of charged particles as tunneling from Reissner-Nordstrom-de Sitter black holes with a global monopole}, %
Phys. Lett. \textbf{B 635} (2006) 151 [hep-th/0511123].

\bibitem{rotating black hole}
J.-Y. Zhang and Z. Zhao, %
{\em Hawking radiation via tunneling from Kerr black holes}, %
Mod. Phys. Lett. \textbf{A 20} (2005) 1673; %
{\em New coordinates for Kerr-Newman black hole radiation}, %
Phys. Lett. \textbf{B 618} (2005) 14; %
\\
S.Z. Yang, %
{\em Kerr-Newman-Kasuya black hole tunnelling radiation}, %
Chin. Phys. Lett. \textbf{22} (2005) 2492; %
\\
J. Zhang and Z. Zhao, %
{\em Charged particles' tunnelling from the Kerr-Newman black hole}, %
gr-qc/0512153.
\\
Q.-Q. Jiang, S.-Q. Wu and X. Cai, %
{\em Hawking radiation as tunneling from the Kerr and
Kerr-Newman black holes}, %
Phys. Rev. \textbf{D 73} (2006) 064003 [hep-th/0512351], erratum ibid. D73 (2006) 069902.
\\
A.J.M. Medved, %
{\em Radiation via tunneling in the charged BTZ black hole}, %
Class. and Quant. Grav. 19 (2002) 589 [hep-th/0110289];
\\
E.C. Vagenas, %
{\em Quantum corrections to the Bekenstein-Hawking entropy of the BTZ black hole
via self-gravitation}, %
Phys. Lett. \textbf{B 533} (2002) 302 [hep-th/0109108].

\bibitem{dimensional reduce}
S.P. Robinson and F. Wilczek, %
{\em Relationship between Hawking Radiation and Gravitational Anomalies}, %
Phys. Rev. Lett. \textbf{95} (2005) 011303 [gr-qc/0502074]; %
\\
S. Iso, H. Umetsu, and F. Wilczek, %
{\em Hawking Radiation from Charged Black Holes via Gauge and Gravitational Anomalies}, %
Phys. Rev. Lett. \textbf{96} (2006) 151302 [hep-th/0602146].

\bibitem{Kerr}
S. Iso, H. Umetsu, and F. Wilczek, %
{\em Anomalies, Hawking radiations, and regularity in rotating black holes}, %
Phys. Rev. \textbf{D 74} (2006) 044017 [hep-th/0606018].

\bibitem{MP}
S. Iso, T. Morita, and H. Umetsu, %
{\em Quantum Anomalies at Horizon and Hawking Radiations in Myers-Perry Black Holes}, %
JHEP \textbf{04} (2007) 068 [hep-th/0612286].

\bibitem{dtunneling}
K. Umetsu, %
{\em Tunneling Mechanism in Kerr-Newman Black Hole and Dimensional Reduction near the Horizon}, %
Phys. Lett. \textbf{B 692} (2010) 61 [arXiv:1007.1823[hep-th]]; %
{\em Hawking Radiation from Kerr-Newman Black Hole and Tunneling Mechanism}, %
arXiv:0907.1420[hep-th].
\\
K. Matsuno and K. Umetsu, %
{\em Hawking radiation as tunneling from squashed Kaluza-Klein black hole}, %
arXiv:1101.2091 [hep-th].
\\
P.-J. Mao, R. Li, L.-Y. Jia, and J.-R. Ren, %
{\em Hawking radiation of Dirac particles
from the Myers¨CPerry black hole}, %
Eur. Phys. J. \textbf{C 71} (2011) 1527.




\bibitem{chow}
B.D. Chowdhury, %
{\em Problems with Tunneling of Thin Shells from Black Holes}, %
Pramana {\bf 70}, 593 (2008) [hep-th/0605197].
\\
E.T. Akhmedov, V. Akhmedova, and D. Singleton, %
{\em Hawking temperature in the tunneling picture}, %
Phys. Lett. {\bf B 642}, 124 (2006) [hep-th/0608098];
\\
E.T. Akhmedov, V. Akhmedova, T. Pilling, and D. Singleton, %
{\em Thermal radiation of various gravitational backgrounds}, %
Int. J. Mod. Phys. \textbf{A 22}, 1705 (2007) [hep-th/0605137]

\bibitem{time}
V. Akhmedova, T. Pilling, A. de Gill, and D. Singleton, %
{\em Comments on anomaly versus WKB/tunneling methods for
calculating Unruh radiation}, %
Phys. Lett. \textbf{B 673} (2009) 227 [arXiv:0808.3413].
\\
T. Pilling, %
{\em Quasi-classical Hawking Temperatures and Black Hole Thermodynamics}, %
arXiv:0809.2701.
\\
E.T. Akhmedov, T. Pilling, and D. Singleton, %
{\em Subtleties in the quasi-classical calculation of Hawking
radiation}, %
Int. J. Mod. Phys. \textbf{D 17} (2009) 2453 [arXiv:0805.2653].
\\
V. Akhmedova, T. Pilling, A. de Gill, and D. Singleton, %
{\em Temporal contribution to gravitational WKB-like calculations}, %
Phys. Lett. \textbf{B 666} (2008) 269 [arXiv:0804.2289].
\\
T. Zhu, J.R. Ren, and D. Singleton, %
{\em Hawking-like radiation as tunneling from the apparent horizon
in a FRW Universe}, %
Int. Mod. Phys. \textbf{D 19} (2010) 159 [arXiv:0902.2542].

\bibitem{Hawking-Cft}
I. Agullo, J.N.-Salas, G.J. Olmo, and L. Parker, %
{\em Hawking Radiation by Kerr Black Holes and Conformal Symmetry}, %
Phys. Rev. Lett. \textbf{105} (2010) 211305 [arXiv:1006.4404[hep-th]].





 \end{thebibliography}
 \end{document}